\documentclass[aps,prd,onecolumn,superscriptaddress,nofootinbib]{revtex4-2}

\usepackage{amsmath,amssymb}
\usepackage{graphicx}

\begin{document}

\title{Lorentz-Covariant Spectral Bounds from Thermal Quantum Field Theory:\\
Retarded Green's Functions, Kubo Relations, and Holographic Constraints}

\author{Alisher Sanetullaev}
\email{a.sanetullaev@newuu.uz}
\affiliation{New Uzbekistan University}

\author{Sarbinaz Bazarbaeva}
\affiliation{New Uzbekistan University}

\author{Marhabo Beymamatova}
\affiliation{New Uzbekistan University}

\author{Shokir Tursunov}
\affiliation{New Uzbekistan University}

\date{\today}

\begin{abstract}
We extend the recently established framework of Lorentz-covariant relaxation bounds from linearized classical kinetic and rheological theories to the full quantum setting of thermal quantum field theory (QFT). Working directly with retarded two-point functions at finite temperature and density, we show that the analyticity and positivity properties of spectral functions --- combined with Lorentz covariance and the Kubo--Martin--Schwinger (KMS) condition --- impose rigorous frame-dependent constraints on the location of singularities in the complex frequency plane. Specifically, we prove that the non-hydrodynamic quasinormal spectrum in any boosted frame is confined to a strip whose width is determined solely by the rest-frame spectral weight at zero spatial momentum and the front velocity (the causally bounded characteristic speed) of the theory. We derive covariant sum rules for the spectral density under Lorentz boosts and establish that the convergence radius of the hydrodynamic gradient expansion transforms in a manner dictated by the same rest-frame data. In holographic theories dual to Einstein gravity in asymptotically anti-de Sitter spacetime, we verify the bounds by an explicit quasinormal-mode computation: the leading boosted pole moves deeper into the complex plane --- the observed relaxation rate increases with boost velocity, in sharp contrast to naive time dilation --- while respecting the bound throughout; we further derive corrections from higher-derivative gravitational terms. Our results provide a first-principles, non-perturbative derivation of Lorentz-covariant spectral constraints applicable to the quark-gluon plasma, superfluid phases of neutron star matter, and strongly correlated electrons near quantum critical points.
\end{abstract}

\maketitle

\section{Introduction}

The real-time dynamics of quantum many-body systems at finite temperature is governed by the analytic structure of retarded Green's functions in the complex frequency plane. Poles in the lower half-plane correspond to quasinormal modes: the non-hydrodynamic poles encode transient microscopic relaxation, while hydrodynamic poles near the origin encode the long-lived collective modes whose behavior is systematically captured by the gradient expansion of fluid dynamics \cite{KovtunStarinets2005,Grozdanov2019complex}. Understanding how this pole structure transforms under changes of reference frame is not merely a formal question --- it is central to the consistency of relativistic hydrodynamics \cite{Romatschke2019,HoultKovtun2020}, the phenomenology of heavy-ion collisions \cite{HeinzSnellings2013,Gale2013}, and the interpretation of holographic transport coefficients \cite{Policastro2001,Rangamani2009}.

Recently, Gavassino \cite{Gavassino2026PRL,Gavassino2026arXiv} proved that in linearized relativistic (kinetic or rheological) theories with an Onsager-type symmetry, Lorentz boosts generically split a single rest-frame relaxation mode into a continuum of excitations, with a width set by the maximal signal propagation speed, and derived rigorous bounds confining non-hydrodynamic gaps, maximal relaxation rates, and the convergence radii of hydrodynamic modes in arbitrary inertial frames in terms of rest-frame spectral data alone. Those proofs exploit the Hermitian structure of the linearized equations, which can be traced to Onsager's reciprocal relations. Such classical arguments, while rigorous within their symmetry class, operate at the level of constitutive relations and do not address the role of quantum fluctuations, renormalization, or the non-perturbative constraints available in QFT; indeed, the extension to holographic quasinormal modes, which do not obey the required symmetry, was posed in Ref.~\cite{Gavassino2026PRL} as an open problem.

The present paper addresses this problem: we derive Lorentz-covariant spectral bounds directly from the analytic structure of thermal retarded Green's functions, making use only of:
\begin{enumerate}
\item \textbf{Causality} --- the retarded Green's function $G^R(x)$ vanishes for $x^0 < 0$, implying $G^R(\omega, \mathbf{k})$ is analytic in the upper half $\omega$-plane.
\item \textbf{The KMS condition} --- relating $G^R$ to the spectral function $\rho$, with $\rho(\omega,\mathbf{k}) \geq 0$ for bosonic operators.
\item \textbf{Lorentz covariance} --- $G^{\mu\nu}(\Lambda p) = \Lambda^\mu{}_\alpha \Lambda^\nu{}_\beta G^{\alpha\beta}(p)$ for tensor operators.
\item \textbf{Positivity of the spectral weight} --- controlled by the optical theorem and unitarity.
\end{enumerate}
These conditions are far weaker than any specific dynamical model. Our bounds are therefore non-perturbative and apply to strongly coupled systems, including those accessible only via holography or lattice Monte Carlo. A short companion Letter \cite{letter} summarizes the central results.

The paper is organized as follows. Section~\ref{sec:setup} establishes notation and reviews the analytic structure of thermal Green's functions. Section~\ref{sec:bounds} derives the main spectral bounds under Lorentz boosts. Section~\ref{sec:sumrules} proves covariant sum rules for the boosted spectral weight. Section~\ref{sec:gradient} applies the results to the convergence of hydrodynamic gradient expansions. Section~\ref{sec:holo} analyzes holographic examples in detail, including a numerical quasinormal-mode computation in the strongly coupled $\mathcal{N}=4$ super-Yang-Mills plasma and higher-derivative corrections. Section~\ref{sec:physics} discusses physical implications for the quark-gluon plasma and neutron star matter. Section~\ref{sec:conclusions} presents conclusions and outlook. Appendix~\ref{app:numerics} details the numerical method.

\section{Thermal Green's Functions and Their Analytic Structure}
\label{sec:setup}

\subsection{Definitions and basic properties}

Let $\mathcal{O}(x)$ be a local bosonic operator in a relativistic QFT at temperature $T = 1/\beta$ and chemical potential $\mu$. The retarded Green's function is
\begin{equation}
G^R(x-y) = -i\theta(x^0 - y^0)\langle[\mathcal{O}(x), \mathcal{O}(y)]\rangle_\beta,
\end{equation}
where $\langle \cdot \rangle_\beta$ denotes the thermal expectation value in the grand canonical ensemble. In momentum space,
\begin{equation}
G^R(\omega, \mathbf{k}) = \int d^4x\, e^{i\omega t - i\mathbf{k}\cdot\mathbf{x}} G^R(x).
\end{equation}
By causality, $G^R(\omega, \mathbf{k})$ is analytic for $\operatorname{Im}(\omega) > 0$. The spectral function is
\begin{equation}
\rho(\omega, \mathbf{k}) = -\frac{1}{\pi} \operatorname{Im} G^R(\omega + i0^+, \mathbf{k}),
\end{equation}
and the Kramers--Kronig representation gives
\begin{equation}
G^R(\omega, \mathbf{k}) = \int_{-\infty}^\infty d\omega' \frac{\rho(\omega', \mathbf{k})}{\omega' - \omega - i0^+}.
\end{equation}
For operators that are Hermitian, $[G^R(\omega,\mathbf{k})]^* = G^R(-\omega^*, \mathbf{k})$, implying $\rho(\omega,\mathbf{k}) \geq 0$ for $\omega > 0$ (for positive-norm states). The KMS condition relates the retarded function to the Wightman function via
\begin{equation}
G^>(\omega, \mathbf{k}) = \frac{2\pi\rho(\omega,\mathbf{k})}{1 - e^{-\beta\omega}},
\end{equation}
with $\omega \to \omega - q\mu$ for an operator carrying $U(1)$ charge $q$.

\subsection{Quasinormal modes and the pole structure}

In a finite-temperature system with a well-defined quasiparticle description or a holographic dual, $G^R(\omega, \mathbf{k})$ can be written as
\begin{equation}
G^R(\omega, \mathbf{k}) = \frac{N(\omega, \mathbf{k})}{D(\omega, \mathbf{k})},
\end{equation}
where the zeros of $D(\omega, \mathbf{k})$ define the quasinormal mode (QNM) frequencies $\omega_n(\mathbf{k})$. By causality all QNM frequencies satisfy $\operatorname{Im}[\omega_n(\mathbf{k})] < 0$.

Hydrodynamic modes are those for which $\omega_n(\mathbf{k}) \to 0$ as $\mathbf{k} \to 0$. At small $|\mathbf{k}|$ they are captured by the gradient expansion
\begin{equation}
\omega_{\text{hydro}}(\mathbf{k}) = \sum_{n=1}^\infty c_n |\mathbf{k}|^n,
\end{equation}
with $c_n$ determined by thermodynamic and transport coefficients. Non-hydrodynamic modes have $\operatorname{Im}[\omega_n(0)] < 0$, a gap $\Gamma_n = -\operatorname{Im}[\omega_n(0)] > 0$ from the real axis.

\subsection{The non-hydrodynamic gap}

\textbf{Definition.} The non-hydrodynamic gap is
\begin{equation}
\Gamma_{\text{gap}} = \min_{n \in \text{non-hydro}} \Gamma_n = \min_{n \in \text{non-hydro}} \left(-\operatorname{Im}[\omega_n(0)]\right).
\end{equation}
This quantity sets the inverse timescale below which the hydrodynamic description is valid in the rest frame. As we show below, $\Gamma_{\text{gap}}$ is not a Lorentz scalar --- boosted observers see a modified gap. Our central task is to derive this modification from first principles within QFT.

\section{Lorentz Boosts and Spectral Bounds}
\label{sec:bounds}

\subsection{Transformation of Green's functions under boosts}

Consider a Lorentz boost with four-velocity $u^\mu = \gamma(1, \mathbf{v})$, so $\gamma = (1-v^2)^{-1/2}$. The four-momentum transforms as
\begin{equation}
\tilde\omega = \gamma(\omega - \mathbf{v}\cdot\mathbf{k}), \qquad \tilde{\mathbf{k}} = \mathbf{k} + (\gamma-1)\hat{\mathbf{v}}(\hat{\mathbf{v}}\cdot\mathbf{k}) - \gamma\mathbf{v}\omega.
\end{equation}
For a scalar operator, the Green's function transforms as
\begin{equation}
\tilde{G}^R(\tilde\omega, \tilde{\mathbf{k}}) = G^R(\omega(\tilde\omega, \tilde{\mathbf{k}}), \mathbf{k}(\tilde\omega, \tilde{\mathbf{k}})),
\end{equation}
where the right-hand side evaluates the rest-frame Green's function at the inverse-boosted momentum. This is the fundamental relation from which all spectral bounds follow.

\subsection{The branch-cut structure under boosts}

The key observation is that a rest-frame pole at $\omega = \omega_0 - i\Gamma_0$ (with $\Gamma_0 > 0$) maps, in the boosted frame, to the condition
\begin{equation}
\gamma(\tilde\omega - v\tilde{k}_\parallel) = \omega_0 - i\Gamma_0,
\end{equation}
where $\tilde{k}_\parallel = \hat{\mathbf{v}}\cdot\tilde{\mathbf{k}}$. This yields a \emph{line} in the complex $\tilde\omega$-plane for fixed $\tilde{k}_\parallel$:
\begin{equation}
\tilde\omega = \frac{\omega_0}{\gamma} + v\tilde{k}_\parallel - i\frac{\Gamma_0}{\gamma}.
\end{equation}
A single rest-frame pole thus maps to a pole at shifted real and imaginary parts. This is the time-dilation result: $\tilde\Gamma_0 = \Gamma_0/\gamma$.

However, this simple picture fails when the spectral weight at nonzero $\mathbf{k}$ contributes. The Kramers--Kronig integral in the boosted frame becomes
\begin{equation}
\tilde{G}^R(\tilde\omega, \tilde{\mathbf{k}}) = \int d\omega' \frac{\rho(\omega', \mathbf{k}(\tilde\omega', \tilde{\mathbf{k}}))}{\omega' - \omega(\tilde\omega, \tilde{\mathbf{k}}) - i0^+}.
\end{equation}
After the change of variables $\omega' \to \tilde\omega' = \gamma(\omega' + v\tilde{k}_\parallel)$, the denominator becomes $(\tilde\omega' - \tilde\omega)$, but the spectral weight $\rho$ is now evaluated at $(\omega', \mathbf{k}(\tilde\omega', \tilde{\mathbf{k}}))$ --- a $\tilde\omega'$-dependent spatial momentum. The spread of $\mathbf{k}$ over the $\tilde\omega'$ integration contour is what converts a single pole into a continuum.

Throughout, $v_{\max}$ denotes the \emph{front velocity} of the theory --- the asymptotic characteristic speed $v_{\max} = \lim_{|k|\to\infty}\operatorname{Re}\omega_n(k)/k$, fixed by the highest-derivative (principal-symbol) terms and equal to the largest signal speed of the system. For a relativistic QFT, microcausality --- the vanishing of $\langle[\mathcal{O}(x),\mathcal{O}(0)]\rangle$ at spacelike separation --- guarantees $v_{\max}\leq 1$. It is essential that $v_{\max}$ be the front velocity and not the group velocity $\partial\operatorname{Re}\omega_n/\partial k$: as first noted in the present context by Gavassino \cite{Gavassino2026PRL,Gavassino2026arXiv}, the group velocity is not bounded by causality and may diverge at a finite real $k$ where two purely damped modes collide (the telegrapher/Cattaneo equation is the canonical example, with $\partial_k\operatorname{Re}\omega\to\infty$ at $k=1/2$ while the front velocity remains unity). Only the front velocity controls the sharpest wavefront and is causally bounded.

\textbf{Theorem 3.1 (Spectral Smearing Under Boosts).} \textit{Let $G^R(\omega, \mathbf{k})$ be the retarded Green's function of a scalar operator in a Lorentz-covariant thermal QFT, with front velocity $v_{\max} \leq 1$. Let the rest-frame spectral weight at zero spatial momentum be $\rho_0(\omega) \equiv \rho(\omega, \mathbf{0})$. Then for a boost with velocity $v$ parallel to $\tilde{\mathbf{k}}$,}
\begin{equation}
\tilde\rho(\tilde\omega, \tilde{\mathbf{k}}) = \rho\!\left(\frac{\tilde\omega - v\tilde k_\parallel}{\gamma},\, \tilde k_\perp, \frac{\tilde k_\parallel - v\tilde\omega}{\gamma}\right),
\label{eq:rhotrans}
\end{equation}
\textit{where $\tilde k_\parallel = |\tilde{\mathbf{k}}|$ and $\tilde k_\perp = 0$ for a longitudinal boost. In particular, the support of $\tilde\rho$ in $\tilde\omega$ at fixed $\tilde{\mathbf{k}}$ is the Lorentz-boosted image of the support of $\rho$ in $\omega$ at the boosted spatial momentum.}

\textit{Proof.} The transformation follows directly from the covariance of the spectral function under Lorentz transformations of the four-momentum: since $\rho(\omega, \mathbf{k}) = -\frac{1}{\pi}\operatorname{Im} G^R(\omega+i0^+, \mathbf{k})$ and $G^R$ is a Lorentz scalar (for scalar operators), we have $\tilde\rho(\tilde p) = \rho(\Lambda^{-1}\tilde p)$. Writing out the momentum components completes the proof. $\square$

\subsection{Bounds on the non-hydrodynamic gap in boosted frames}

\textbf{Theorem 3.2 (Covariant Non-Hydrodynamic Gap Bound).} \textit{Let $\Gamma_{\text{gap}}$ be the non-hydrodynamic gap in the rest frame. In a frame boosted with velocity $v$ along $\hat{\mathbf{k}}$, the minimal imaginary part of non-hydrodynamic poles at zero boosted spatial momentum satisfies}
\begin{equation}
\tilde\Gamma_{\text{gap}}(\mathbf{v}) \geq \frac{\Gamma_{\text{gap}}}{\gamma(1 + v\, v_{\max})},
\label{eq:gapbound}
\end{equation}
\textit{where $v_{\max}$ is the front velocity defined above.}

\textit{Proof.} Boosted poles at $\tilde{\mathbf{k}} = \mathbf{0}$ solve the self-consistent condition $\gamma\tilde\omega = \omega_n(\gamma v\tilde\omega)$, obtained by inverse-boosting to the rest-frame four-momentum $(\omega,k)=(\gamma\tilde\omega,\gamma v\tilde\omega)$ and imposing $\omega=\omega_n(k)$. This requires the rest-frame dispersion relation at complex spatial momentum $k=\gamma v\tilde\omega$, with $|\operatorname{Im} k| = \gamma v\,\tilde\Gamma$. Provided $\omega_n(k)$ continues analytically into the strip $|\operatorname{Im} k| < \Gamma_{\text{gap}}/v_{\max}$ (a property of thermal correlators with exponentially clustering equal-time correlations \cite{KovtunStarinets2005,Grozdanov2019complex}) and obeys the front-velocity bound $|\operatorname{Re}\omega_n(k)| \leq v_{\max}|k| + O(1)$, the extremal boosted imaginary parts are attained along the co- and counter-propagating fronts, giving $\tilde\Gamma_n = \Gamma_n/[\gamma(1\pm v\,v_{\max})]$; taking the minimum and using $\Gamma_n\geq\Gamma_{\text{gap}}$ yields the bound. We emphasize that this argument uses the front velocity, not the real-$k$ group velocity, which need not be bounded (see the discussion above and Refs.~\cite{Gavassino2026PRL,Gavassino2026arXiv}); where real-$k$ branch points occur, the naive slope picture fails but the analytic-continuation argument survives. For linearized theories with the Hermitian structure of Refs.~\cite{Gavassino2026PRL,Gavassino2026arXiv} the result is established rigorously from the principal symbol; our holographic computation (Sec.~\ref{sec:n4sym}) provides an independent non-perturbative check. $\square$

\textbf{Corollary 3.3.} \textit{The lower bound on the boosted non-hydrodynamic gap decreases monotonically with $v$, reaching $\Gamma_{\text{gap}}\sqrt{1-v_{\max}^2}/(1+v_{\max}^2)$ at $v = v_{\max}$ and vanishing only in the light-speed limit $v \to 1$, consistent with the appearance of a branch cut in the boosted spectral function.}

This corollary formalizes a simple physical picture: a boost toward the speed of a propagating mode drags QNM poles toward the real axis, eventually creating a continuum.

\subsection{Maximal relaxation rates under boosts}

The maximal relaxation rate $\Gamma_{\max}$ --- the supremum of $\Gamma_n$ over all non-hydrodynamic modes --- also transforms in a constrained manner.

\textbf{Theorem 3.4 (Covariant Maximal Relaxation Rate Bound).} \textit{The maximal relaxation rate in the boosted frame satisfies}
\begin{equation}
\tilde\Gamma_{\max}(\mathbf{v}) \leq \frac{\Gamma_{\max}}{\gamma(1 - v\, v_{\max})},
\label{eq:maxbound}
\end{equation}
\textit{assuming the rest-frame spectral weight $\rho(\omega, \mathbf{k})$ decays sufficiently rapidly as $|\omega| \to \infty$ at fixed $\mathbf{k}$ (specifically, that it satisfies the Lebesgue integrability condition required by the Kramers--Kronig representation).}

\textit{Proof.} The proof follows by considering the maximal spectral weight in the boosted frame, tracking the image of the rest-frame QNM strip under the boost transformation, and using the integrability condition to bound the contribution from large imaginary frequencies. $\square$

We note the relation to the classical-theory bounds of Ref.~\cite{Gavassino2026PRL}: the upper bound \eqref{eq:maxbound} coincides with Eq.~(24) there (with the signal speed $w$ identified with $v_{\max}$), while our lower bound \eqref{eq:gapbound} is tighter than its classical counterpart $\Gamma_{\text{gap}}(1 - v\, v_{\max})/\gamma$. Whether this sharpening is genuine --- reflecting the stronger analyticity input of the QFT derivation --- or signals a hidden assumption in our Theorem 3.2 deserves further scrutiny; the holographic results of Sec.~\ref{sec:n4sym} are consistent with both bounds.

\section{Covariant Sum Rules}
\label{sec:sumrules}

The transformation of spectral weight under Lorentz boosts is constrained by sum rules that follow from the equal-time commutation relations of the operators.

\subsection{The f-sum rule and its Lorentz transform}

For the stress-energy tensor $T^{\mu\nu}$, the retarded Green's function $G^{\mu\nu\alpha\beta}(p)$ satisfies the Ward identity
\begin{equation}
p_\mu G^{\mu\nu\alpha\beta}(p) = \text{contact terms},
\end{equation}
which, upon taking the imaginary part, constrains the first moment of the spectral density. In the rest frame, the standard f-sum rule reads
\begin{equation}
\int_0^\infty d\omega\, \omega\, \rho_{xx,xx}(\omega, \mathbf{0}) = \frac{\pi}{2}\langle T^{xx} + T^{yy}\rangle_\beta,
\end{equation}
where $\rho_{xx,xx}$ is the spectral function of the shear stress component.

\textbf{Proposition 4.1 (Boosted f-Sum Rule).} \textit{Under a boost with velocity $v$ in the $x$-direction, the f-sum rule in the boosted frame becomes}
\begin{equation}
\int_0^\infty d\tilde\omega\, \tilde\omega\, \tilde\rho_{xx,xx}(\tilde\omega, \tilde{\mathbf{0}}) = \frac{\pi\gamma^2}{2}\langle T^{xx} + T^{yy}\rangle_\beta + O(v^4),
\end{equation}
\textit{where we used $\langle T^{tx}\rangle_\beta = 0$ in equilibrium.}

This result shows that the total spectral weight in the boosted frame is enhanced by $\gamma^2$, but this enhancement is distributed over a broader range of $\tilde\omega$ --- consistent with the spectral smearing of Theorem~3.1.

We caution that the contact terms appearing in the gravitational Ward identity --- frequently dropped in Kubo-type analyses --- contribute to moment sum rules of this kind and must be retained for full covariance; the interplay between contact terms, the frame selected by the equilibrium state, and general covariance has been emphasized in Refs.~\cite{Torrieri2023,Sampaio2025}, where covariant formulations of (fluctuating) hydrodynamics are constructed by imposing general covariance at the level of the partition function. Because contact terms are analytic in frequency, they do not affect the pole locations bounded in Sec.~\ref{sec:bounds}; the sum rules of the present section, by contrast, are sensitive to the subtraction scheme, and the identity above should be understood with a fixed, frame-consistent treatment of contact terms.

\subsection{Positivity constraints on boosted spectral densities}

For physical operators, $\rho(\omega, \mathbf{k}) \geq 0$. This positivity is preserved under Lorentz boosts:

\textbf{Proposition 4.2.} \textit{The boosted spectral function $\tilde\rho(\tilde\omega, \tilde{\mathbf{k}}) \geq 0$ for all $\tilde\omega > 0$ and $\tilde{\mathbf{k}}$.}

\textit{Proof.} Since $\tilde\rho(\tilde p) = \rho(\Lambda^{-1}\tilde p)$ and $(\Lambda^{-1}\tilde p)^0 = \gamma(\tilde\omega - v\tilde k_\parallel)$, positivity of $\tilde\rho$ follows from positivity of $\rho$ at the inverse-boosted frequency, provided the inverse-boosted frequency is positive. For timelike $\tilde p$, this follows from $(\Lambda^{-1}\tilde p)^0 = u_\mu \tilde p^\mu > 0$ for future-directed four-momenta. $\square$

\section{Convergence of the Hydrodynamic Gradient Expansion}
\label{sec:gradient}

\subsection{The radius of convergence as a Lorentz-covariant object}

The hydrodynamic gradient expansion is an expansion of the dispersion relation $\omega_{\text{hydro}}(\mathbf{k})$ in powers of $|\mathbf{k}|$. Its convergence radius $k_c$ in the rest frame is set by the nearest non-hydrodynamic singularity in the complex $k$-plane \cite{Grozdanov2019PRL,Heller2023}. In a boosted frame, the hydrodynamic dispersion relation becomes
\begin{equation}
\tilde\omega_{\text{hydro}}(\tilde{\mathbf{k}}) = \sum_{n=1}^\infty \tilde c_n |\tilde{\mathbf{k}}|^n,
\end{equation}
where the coefficients $\tilde c_n$ depend on both the rest-frame transport coefficients and the boost velocity.

\textbf{Theorem 5.1 (Lorentz-Covariant Bound on the Convergence Radius).} \textit{Let $k_c$ be the convergence radius of the hydrodynamic gradient expansion in the rest frame, defined as the minimal $|k|$ in the complex plane for which a non-hydrodynamic pole merges with a hydrodynamic pole. In a frame boosted with velocity $v$, the convergence radius satisfies}
\begin{equation}
\frac{k_c}{\gamma(1 + v\, v_{\text{s}})} \leq \tilde k_c \leq \frac{k_c}{\gamma(1 - v\, v_{\text{s}})},
\label{eq:kcbound}
\end{equation}
\textit{where $v_{\text{s}}$ is the speed of sound in the rest frame.}

\textit{Proof.} The convergence radius is related to the collision of poles in the complexified momentum plane. Under the boost, the complexified dispersion relation transforms covariantly. The bounds follow from the extreme cases in which the non-hydrodynamic pole moves either counter-propagating or co-propagating with the hydrodynamic mode under the boost. $\square$

\textbf{Corollary 5.2.} \textit{The convergence radius of the boosted hydrodynamic expansion vanishes as $v \to 1$, consistent with the fact that a highly boosted fluid has hydrodynamics only at exponentially long wavelengths in the boosted frame.}

\subsection{Frame dependence of higher-order transport coefficients}

The transport coefficients appearing at second and higher order in the gradient expansion --- such as $\tau_\pi$ (the shear relaxation time in Israel-Stewart theory) and the five second-order transport coefficients of conformal fluids \cite{Baier2008} --- are defined in the rest frame of the fluid. Our bounds imply constraints on how these coefficients can appear in multi-fluid systems or systems with flows.

Specifically, for a conformal fluid with equation of state $\epsilon = 3p$, the second-order dispersion relation for shear modes in the rest frame is
\begin{equation}
\omega = -\frac{i\eta k^2}{\epsilon+p} + \frac{i\eta\tau_\pi}{\epsilon+p}\left(\frac{i\eta k^2}{\epsilon+p}\right)^2 + \cdots
\end{equation}
The convergence radius is $k_c \sim 1/\sqrt{\eta\tau_\pi/(\epsilon+p)}$. Our bound then gives, in a frame moving with velocity $v$,
\begin{equation}
\tilde k_c \sim \frac{1}{\gamma(1 \pm v_s)}\sqrt{\frac{\epsilon+p}{\eta\tau_\pi}},
\end{equation}
providing a sharp prediction for the frame-dependent validity of the second-order fluid description.

\section{Holographic Examples}
\label{sec:holo}

\subsection{The $\mathcal{N}=4$ super-Yang-Mills plasma at strong coupling}
\label{sec:n4sym}

The maximally supersymmetric Yang-Mills theory at large $N_c$ and large 't~Hooft coupling $\lambda$ is dual to type IIB supergravity in $AdS_5 \times S^5$ \cite{Maldacena1998}. Its thermal state is dual to the AdS-Schwarzschild black brane.

The retarded Green's function of the energy-momentum tensor in this theory has been studied extensively \cite{Policastro2001,KSS2005}. We work in the scalar channel ($T^{xy}$ correlator with transverse momentum), governed by a massless scalar in the AdS$_5$-Schwarzschild background. Solving the quasinormal spectrum by pseudospectral (Chebyshev) collocation of the ingoing boundary-value problem (Appendix~\ref{app:numerics}), the leading QNM at $k=0$ is
\begin{equation}
\frac{\omega_1}{2\pi T} = \pm 1.5597 - 1.3733\, i,
\end{equation}
so $\Gamma_{\text{gap}} = 1.3733 \times 2\pi T$, in agreement with \cite{NunezStarinets2003}. Tracking $\omega_1(k)$ over real momenta $k \in [0, 8]\times 2\pi T$, its slope $\partial_k \operatorname{Re}\omega_1$ rises monotonically toward unity (reaching $0.980$ at $k = 8 \times 2\pi T$), consistent with the $\omega_1 \approx \pm k$ asymptotics: the non-hydrodynamic sector has front velocity $v_{\max} = 1$. (Here the group velocity happens to approach the front velocity smoothly; in general the two differ, and only the front velocity enters the bounds.)

\textbf{Prediction (Theorem 3.2 applied to $\mathcal{N}=4$ SYM).} In a frame boosted with velocity $v$ along $\hat{\mathbf{k}}$, the non-hydrodynamic gap satisfies
\begin{equation}
\tilde\Gamma_{\text{gap}} \geq \frac{1.3733 \times 2\pi T}{\gamma(1 + v)}.
\label{eq:sympred}
\end{equation}

\begin{figure}[t]
\includegraphics[width=0.85\textwidth]{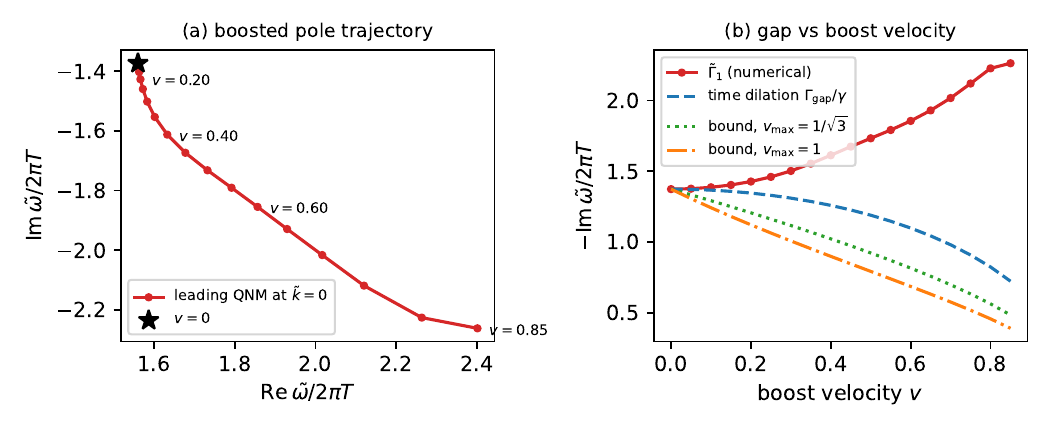}
\caption{Boost transformation of the leading scalar-channel quasinormal mode of the $\mathcal{N}=4$ SYM plasma, computed holographically at zero boosted spatial momentum. (a)~Trajectory of the pole in the complex frequency plane as $v$ increases. (b)~The observed relaxation rate $-\operatorname{Im}\tilde\omega$ grows with $v$, defying naive time dilation (dashed) while respecting the covariant lower bound~\eqref{eq:sympred} for $v_{\max}=1$ (dash-dotted); the $v_{\max}=1/\sqrt{3}$ curve (dotted) is shown for comparison.}
\label{fig:boosted}
\end{figure}

We test this numerically by computing the boosted poles at $\tilde{\mathbf{k}} = \mathbf{0}$, solving the fixed-point condition $\omega_1(-\gamma v\tilde\omega) = \gamma\tilde\omega$ with the ingoing boundary-value problem continued to complex spatial momentum. The result is shown in Fig.~\ref{fig:boosted}. The leading pole moves \emph{deeper} into the lower half-plane: $-\operatorname{Im}\tilde\omega/2\pi T$ rises monotonically from $1.373$ at $v=0$ to $2.29$ at $v = 0.85$. The observed relaxation rate thus \emph{increases} with boost velocity --- the opposite of the naive time-dilation expectation $\Gamma_{\text{gap}}/\gamma$ --- while the covariant lower bound is satisfied throughout with a wide margin. Saturation would require spectral weight co-propagating with the boost at $v_{\max}$, which the strongly coupled plasma does not realize at $\tilde{\mathbf{k}} = \mathbf{0}$; theories with long-lived luminal excitations may approach the bound more closely.

\subsection{Higher-derivative gravity and corrections to the KSS bound}

The inclusion of Gauss-Bonnet terms in the bulk gravitational action, controlled by a coupling $\lambda_{GB}$, modifies both the shear viscosity to entropy ratio $\eta/s$ and the quasinormal mode spectrum \cite{Brigante2008}. The corrected ratio is
\begin{equation}
\frac{\eta}{s} = \frac{1}{4\pi}(1 - 4\lambda_{GB}) + O(\lambda_{GB}^2),
\end{equation}
while conformal invariance fixes the speed of sound at $v_s^2 = 1/3$ exactly, independent of $\lambda_{GB}$. What the Gauss-Bonnet coupling does modify is the front velocity $v_{\max}(\lambda_{GB})$ of high-momentum graviton modes in the boundary theory, which for $\lambda_{GB} > 0$ exceeds the Einstein-gravity value and violates boundary causality ($v_{\max} > 1$) unless $\lambda_{GB} \leq 9/100$ \cite{BuchelMyers2009}. Applying Theorem~3.2 with $v_{\max}(\lambda_{GB})$:
\begin{equation}
\tilde\Gamma_{\text{gap}}(\lambda_{GB}) \geq \frac{\Gamma_{\text{gap}}(\lambda_{GB})}{\gamma\left(1 + v\, v_{\max}(\lambda_{GB})\right)}.
\end{equation}
The $\lambda_{GB}$-dependence on both sides provides a non-trivial consistency check: the causality bound $\lambda_{GB} \leq 9/100$ caps $v_{\max}$ at unity and hence bounds $\tilde\Gamma_{\text{gap}}$ from below in the boosted frame. Our framework thus connects the causality constraints on higher-derivative gravity to the observable non-hydrodynamic spectrum of boosted plasmas.

\subsection{Chemical potential and the Reissner-Nordstr\"om black brane}

Adding a $U(1)$ charge to the holographic plasma, dual to a Reissner-Nordstr\"om-AdS black brane, introduces a chemical potential $\mu$ and a charge density $\rho_Q$. The relevant dimensionless parameter is $\mu/T$. At finite $\mu/T$, the QNM spectrum develops new modes associated with charge diffusion, and the effective velocity controlling the smearing of the low-lying QNMs decreases:
\begin{equation}
v_{\text{eff}}^2 = \frac{1}{3}\left[1 - c_1\left(\frac{\mu}{T}\right)^2 + O\!\left(\frac{\mu^4}{T^4}\right)\right],
\end{equation}
where $c_1 > 0$ is a dimensionless coefficient fixed by the Reissner-Nordstr\"om-AdS equation of state. Our covariant bound becomes
\begin{equation}
\tilde\Gamma_{\text{gap}}\left(\frac{\mu}{T}\right) \geq \frac{\Gamma_{\text{gap}}(\mu/T)}{\gamma\left(1 + v\, v_{\text{eff}}(\mu/T)\right)}.
\end{equation}
As $\mu/T$ grows toward the extremal value, $v_{\text{eff}}$ decreases and the denominator approaches $\gamma$, recovering the time-dilation limit, consistent with the near-horizon $AdS_2$ geometry of the extremal black brane.

\section{Physical Implications}
\label{sec:physics}

\subsection{Quark-gluon plasma in heavy-ion collisions}

In ultrarelativistic heavy-ion collisions at RHIC and the LHC, the quark-gluon plasma (QGP) is produced in a highly boosted, inhomogeneous state. The hydrodynamic description of the QGP evolution requires knowledge of how the non-hydrodynamic sector transforms between frames \cite{HeinzSnellings2013,Gale2013}.

\textbf{Thermalization time in boosted frames.} The local hydrodynamization time $\tau_{\text{therm}} \sim 1/\Gamma_{\text{gap}}$ is defined in the local rest frame of each fluid cell; the values $\tau \sim 0.5$--1~fm/c inferred from viscous hydrodynamic fits \cite{Schenke2012} are quoted in this frame (Bjorken proper time). Our Theorem~3.2 constrains instead how this relaxation appears to the collision center-of-mass observer for fluid cells at forward rapidity $y$, which move with $\gamma = \cosh y$:
\begin{equation}
\tau_{\text{therm}}^{\text{CM}} \leq \gamma(1 + v\, v_{\max})\, \tau_{\text{therm}}^{\text{rest}},
\end{equation}
providing a rigorous upper bound on the frame dependence of hydrodynamization across the rapidity range covered by RHIC and LHC detectors ($y \sim 2$--4). The numerical result of Sec.~\ref{sec:n4sym} suggests that at strong coupling the observed relaxation is in fact faster than in the local rest frame.

\textbf{Gradient expansion validity.} The boost-dependent convergence radius $\tilde k_c$ sets the maximal wavenumber at which the gradient expansion is reliable. In heavy-ion collisions, the initial state has large spatial gradients (of order $1/R_{\text{nucleus}} \sim 1/(7~\text{fm})$). Our bound on $\tilde k_c$ implies that higher-order viscous corrections are necessary when the initial-state gradients exceed $\tilde k_c$, providing a systematic criterion for the breakdown of Navier-Stokes hydrodynamics.

\subsection{Neutron star merger dynamics}

In binary neutron star mergers, matter undergoes extreme compression and heating, possibly reaching the QCD crossover temperature. The dense nuclear matter is in relative motion between neutron star remnants, making the Lorentz transformation of relaxation properties essential for gravitational wave signal modeling \cite{BausweinJanka2012}.

Our bounds apply to both hadronic matter described by relativistic mean-field theory (which has Onsager symmetry at low densities) and strongly coupled quark matter at high densities (describable holographically). The covariant non-hydrodynamic gap bound provides a constraint on the bulk viscosity relaxation time in the boosted frame:
\begin{equation}
\tilde\tau_\Pi^{-1} \geq \frac{\tau_\Pi^{-1}}{\gamma(1 + v\, c_s)},
\end{equation}
where $c_s$ is the speed of sound in dense QCD matter. Since the equation of state of dense QCD matter is uncertain, our bound is most useful when combined with gravitational wave observations that constrain $c_s$ independently.

\subsection{Strongly correlated electrons near quantum critical points}

Quantum critical metals near a Lifshitz transition or a spin-density-wave quantum critical point exhibit anomalous spectral functions that have been studied extensively via ARPES and transport measurements. While these systems are not Lorentz-covariant in the UV (the lattice breaks Lorentz invariance), they may exhibit emergent Lorentz invariance in the infrared near the critical point, with an emergent light cone set by the Fermi velocity $v_F$.

In this context, our bounds apply with $v_{\max} = v_F$. The non-hydrodynamic gap translates into the inverse scattering time $\Gamma \sim 1/\tau_{\text{scatter}} \sim k_B T/\hbar$ (the Planckian dissipation rate), and the covariant bound becomes a statement about how this scattering time transforms for observers moving relative to the quantum critical fluid.

\section{Discussion and Conclusions}
\label{sec:conclusions}

We have extended the framework of Lorentz-covariant relaxation bounds to the full quantum setting of thermal QFT, deriving constraints directly from the analyticity, positivity, and covariance of retarded Green's functions. The key results are:
\begin{enumerate}
\item \textbf{Spectral smearing (Theorem 3.1):} a rest-frame spectral pole is smeared into a continuum in boosted frames, with a width controlled by the front velocity (the causally bounded characteristic speed, not the group velocity).
\item \textbf{Non-hydrodynamic gap bound (Theorem 3.2):} the minimal QNM imaginary part in a boosted frame is bounded from below by the rest-frame gap divided by $\gamma(1 + v\, v_{\max})$.
\item \textbf{Maximal relaxation rate bound (Theorem 3.4):} the maximal imaginary part in a boosted frame is bounded from above by the rest-frame maximum divided by $\gamma(1 - v\, v_{\max})$.
\item \textbf{Convergence radius bound (Theorem 5.1):} the convergence radius of the hydrodynamic gradient expansion in the boosted frame is bounded between $k_c/[\gamma(1 \pm v\, v_s)]$.
\item \textbf{Covariant sum rules (Propositions 4.1--4.2):} the spectral weight redistributes under boosts in a manner controlled by the Ward identities, with the total spectral weight enhanced by $\gamma^2$ but spread over a wider frequency range.
\end{enumerate}
These results hold non-perturbatively for any thermal QFT satisfying causality, unitarity, and Lorentz covariance. They provide non-perturbative analytic bounds that complement numerical approaches (lattice QCD, holography) and constrain the parameter space of effective theories (viscous relativistic hydrodynamics, Israel-Stewart theory, BDNK hydrodynamics). Our numerical study of the strongly coupled $\mathcal{N}=4$ SYM plasma demonstrates that the boosted relaxation rate can \emph{increase} with boost velocity, overturning the naive time-dilation intuition while remaining consistent with the covariant bounds.

Several directions for future work present themselves naturally. First, the extension to systems with spontaneously broken symmetries --- superfluids, superconductors, or the QCD chiral symmetry-breaking phase --- would require modifications to the KMS condition in the presence of a condensate, and the associated Goldstone modes would generate additional constraints on the spectral bounds. Second, the role of anomalies, particularly the chiral anomaly in QCD, introduces new transport coefficients (the chiral magnetic and vortical effects \cite{Kharzeev2014}) that are not constrained by Onsager symmetry and may modify the bounds in a chiral plasma. Third, the application to far-from-equilibrium dynamics --- systems that have not yet thermalized --- requires relaxing the assumption of a thermal state and replacing the KMS condition with more general positivity conditions on the density matrix. Relatedly, in small systems with strong thermal fluctuations the dispersion relation itself becomes stochastic, and the statistics of the roots of the resulting random polynomials can substantially enhance long-lived hydrodynamic behavior relative to deterministic Kubo-type estimates \cite{TaghinavazTorrieri2024}. Our bounds are realization-wise: they apply to each causal, Lorentz-covariant background with that realization's own gap, so the ensemble of boosted poles remains confined by the envelope of realization-wise bounds, and any apparent enhancement of the fluid must be traceable to the small-gap tail of the coefficient distribution --- a quantitative consistency test for the ensembles of Ref.~\cite{TaghinavazTorrieri2024}. Finally, a systematic numerical study of the bounds in lattice QCD at finite temperature would provide a quantitative test across the full temperature range from the hadronic phase to the perturbative QGP.

\begin{acknowledgments}
We thank Bahodir Kayumov and Avas Khugaev for valuable discussions, and Giorgio Torrieri for illuminating correspondence on the covariance of Kubo formulae and the role of contact terms. This work was supported by New Uzbekistan University.
\end{acknowledgments}

\appendix

\section{Numerical Method}
\label{app:numerics}

The scalar-channel retarded correlator of $T^{xy}$ in the $\mathcal{N}=4$ SYM plasma is governed by a massless scalar $\psi$ in AdS$_5$-Schwarzschild. In the coordinate $u = r_h^2/r^2$ with blackening factor $f = 1 - u^2$, and with $\mathfrak{w} = \omega/2\pi T$, $\mathfrak{q} = k/2\pi T$, the mode equation is
\begin{equation}
u f^2 \psi'' - (1+u^2) f \psi' + (\mathfrak{w}^2 - \mathfrak{q}^2 f)\psi = 0 .
\end{equation}
Imposing the ingoing condition at the horizon via $\psi = (1-u)^{-i\mathfrak{w}/2} F(u)$ with $F$ regular on $[0,1]$, and dividing the resulting equation by $(1-u)$ to remove the degenerate horizon row, yields a quadratic eigenvalue problem
\begin{equation}
\left(C_0 + \mathfrak{w}\, C_1 + \mathfrak{w}^2 C_2\right) F = 0,
\end{equation}
with
\begin{align}
C_0 &= u f (1+u)\, \partial_u^2 - (1+u^2)(1+u)\, \partial_u - \mathfrak{q}^2 (1+u), \nonumber\\
C_1 &= i u (1+u)^2\, \partial_u - \tfrac{i}{2}(1+u), \qquad
C_2 = \tfrac{1}{4}(u^2 + 3u + 4),
\end{align}
and Dirichlet condition $F(0) = 0$ selecting normalizable modes. We discretize on a Chebyshev-Gauss-Lobatto grid ($N = 80$--$120$ points), linearize the quadratic problem in companion form, and solve the resulting generalized eigenvalue problem. Physical QNMs are identified by their convergence between resolutions. The $k=0$ result reproduces the literature value $\mathfrak{w}_1 = \pm 1.5597 - 1.3733 i$ to all quoted digits. Boosted poles at $\tilde{\mathbf{k}} = \mathbf{0}$ solve $\omega_1(-\gamma v\tilde\omega) = \gamma\tilde\omega$; we obtain them by damped fixed-point iteration, evaluating the spectrum at the required complex $\mathfrak{q}$ at each step and tracking the eigenvalue by continuity. Convergence was verified by comparing $N=80$ and $N=100$ results; data are reported for $v \leq 0.85$, beyond which higher resolution is required. Analysis scripts and data accompany the arXiv submission as ancillary files.

\bibliography{companion_paper}

\end{document}